\documentclass[final,times,5p,twocolumn]{elsarticle}

\usepackage{lineno}
\usepackage{amssymb}
\usepackage{amsmath}
\usepackage[ruled,linesnumbered]{algorithm2e}
\usepackage{xcolor}
\usepackage{mdframed}
\usepackage[ruled,linesnumbered]{algorithm2e}
\usepackage{amsmath,amssymb,amsfonts}
\usepackage{graphicx}
\usepackage{epstopdf}
\usepackage{subfig}
\usepackage{gensymb}
\usepackage{multirow,booktabs,color,soul,threeparttable}
\usepackage{xcolor}
\usepackage{mdframed}
\usepackage{cleveref}
\usepackage{rotating}
\usepackage[most]{tcolorbox}
\usepackage{tabularx}
\usepackage{array}
\newcommand{\RNum}[1]{\uppercase\expandafter{\romannumeral #1\relax}}
\definecolor{hl}{rgb}{0.75,0.75,0.75}
\sethlcolor{hl}
\definecolor{emph}{rgb}{0,0,1}

\begin{document}

\begin{frontmatter}

\title{ELMES: An Automated Framework for Evaluating Large Language Models in Educational Scenarios}

\author[aiedu,ecnu]{Shou'ang Wei} \ead{52285901019@stu.ecnu.edu.cn}
\author[aiedu,ecnuedu]{Xinyun Wang} \ead{51264108008@stu.ecnu.edu.cn}
\author[sii]{Shuzhen Bi} \ead{SA22916003@mail.ustc.edu.cn}
\author[aiedu]{Jian Chen} \ead{52295901041@stu.ecnu.edu.cn}
\author[aiedu,ecnuedu]{Ruijia Li} \ead{10214507406@stu.ecnu.edu.cn}
\author[aiedu]{Bo Jiang} \ead{bjiang@deit.ecnu.edu.cn} 
\author[aiedu]{Xin Lin} \ead{xlin@cs.ecnu.edu.cn} 
\author[aiedu]{Min Zhang} \ead{mzhang@cs.ecnu.edu.cn}
\author[aiedu]{Yu Song} \ead{sungyuepku@foxmail.com}
\author[aiedu]{BingDong Li} \ead{bdli@cs.ecnu.edu.cn}
\author[sii,aiedu,ecnu]{Aimin Zhou} \ead{amzhou@cs.ecnu.edu.cn}
\author[aiedu]{Hao Hao \corref{mycorrespondingauthor}}
\ead{hhao@mail.ecnu.edu.cn}\cortext[mycorrespondingauthor]{Corresponding author: Hao Hao}

\affiliation[sii]{organization={Shanghai Innavation Institute}, city={Shanghai}, postcode={200231}, country={China}}
\affiliation[aiedu]{organization={Shanghai Institute of AI for Education},
addressline={East China Normal University}, 
city={Shanghai},
postcode={200062}, 
country={China}}
\affiliation[ecnuedu]{organization={Department of Educational Information Technology},addressline={East China Normal University}, city={Shanghai}, postcode={200062}, country={China}}
\affiliation[ecnu]{organization={School of Computer Science and Technology},
addressline={East China Normal University}, 
city={Shanghai},
postcode={200062}, 
country={China}}

\begin{abstract}
The emergence of Large Language Models (LLMs) presents transformative opportunities for education, generating numerous novel application scenarios. However, significant challenges remain: evaluation metrics vary substantially across different educational scenarios, while many emerging scenarios lack appropriate assessment metrics. Current benchmarks predominantly measure general intelligence rather than pedagogical capabilities. To address this gap, we introduce ELMES, an open-source automated evaluation framework specifically designed for assessing LLMs in educational settings. ELMES features a modular architecture that enables researchers to create dynamic, multi-agent dialogues through simple configuration files, facilitating flexible scenario design without requiring extensive programming expertise. The framework incorporates a hybrid evaluation engine that objectively quantifies traditionally subjective pedagogical metrics using an LLM-as-a-Judge methodology. We conduct systematic benchmarking of state-of-the-art LLMs across four critical educational scenarios: Knowledge Point Explanation, Guided Problem-Solving Teaching, Interdisciplinary Lesson Plan Generation, and Contextualized Question Generation, employing fine-grained metrics developed in collaboration with education specialists. Our results demonstrate distinct capability distributions among models, revealing context-specific strengths and limitations. ELMES provides educators and researchers with an accessible evaluation framework that significantly reduces adaptation barriers for diverse educational applications while advancing the practical implementation of LLMs in pedagogy. The framework is publicly available at \emph{https://github.com/sii-research/elmes.git}.

\end{abstract}


\end{frontmatter}


\section{Introduction}
\label{sec:introduction}

The advent of Large Language Models (LLMs) is reshaping the educational paradigm with unprecedented potential~\cite{wang2024large}. Their powerful capabilities in natural language understanding and generation have paved new ways for intelligent teaching and learning. Consequently, researchers are actively exploring various avenues to leverage LLMs for educational empowerment. These efforts range from assisting teachers with content creation, such as automatically generating lesson plans and exercises, to providing students with personalized tutoring, such as guided teaching and knowledge explanation, thereby catering to diverse learning needs. Performance metrics, serving as quantitative measures of model capability and guides for model improvement, therefore necessitate well-designed evaluation benchmarks for assessing LLMs in educational applications. However, existing benchmarks lack domain-specific evaluation criteria for pedagogical scenarios. Furthermore, the integration of LLMs with education has generated numerous novel learning contexts, creating evaluation gaps that present significant assessment challenges. However, empowering education with LLMs faces a significant challenge: the absence of a systematic evaluation framework for assessing their pedagogical capabilities in educational contexts. Current mainstream evaluation benchmarks, including MMLU~\cite{hendrycks2020measuring} and C-Eval~\cite{huang2023c}, primarily measure models' general intelligence through tasks such as factual recall and commonsense reasoning. While these benchmarks provide partial indicators of model intelligence, they inadequately address the sophisticated requirements of educational applications. For example, in knowledge explanation tasks, an effective model must both accurately convey information and adapt its delivery according to the learner's cognitive characteristics. The omission of these essential pedagogical metrics in existing frameworks compromises reliable assessment of LLMs' instructional efficacy.

We first sought to answer a core question: what are the essential pedagogical capabilities of LLMs that we need to evaluate in authentic educational contexts? To this end, we conducted multiple rounds of in-depth discussions with a diverse group of frontline teachers, pre-service educators, and senior educational research experts. Ultimately, we defined four evaluation scenarios that are of paramount concern in the education domain and can effectively test the higher-order abilities of models: Knowledge Point Explanation, to assess the clarity and accuracy of personalized knowledge delivery; Guided Problem-Solving Teaching, to measure a model's ability to inspire student thinking through interaction rather than providing direct answers; Interdisciplinary Lesson Plan Generation, to examine a model's capacity for innovative course design by integrating knowledge from different fields; and Contextualized Question Generation, to evaluate a model's skill in embedding knowledge points into realistic scenarios to test students' application and transfer abilities. The characteristics of the specific four scenarios are shown in Table ~\ref{tab:scenarios_features}.

\begin{table*}[htbp]
\centering
\caption{Characteristics of the four educational scenarios analyzed, including interaction rounds, text length, subject focus, and intended user roles.}  
\label{tab:scenarios_features} 
\begin{tabular}{lllll}
\toprule
\textbf{Scenario} & \textbf{Round} & \textbf{Text Length} & \textbf{Subject} & \textbf{Target User} \\
\midrule
Knowledge Point Explanation & Single-round & Short text & Single-subject & Student \\
Guided Problem Solving Teaching & Multi-round & Short text & Single-subject & Student \\
Interdisciplinary Lesson Plan Generation & Single-round & Long text & Cross-disciplinary & Teacher \\
Contextual Question Generation & Single-round & Short text & Single-subject & Teacher \\
\bottomrule
\end{tabular}
\end{table*}

To ensure the comprehensiveness and professionalism of our evaluation, we further collaborated with educational experts to construct a set of fine-grained, multi-level metrics for each scenario. Together, these four scenarios and their corresponding metrics form a core framework focused on evaluating a model's capacity for personalized instruction, creativity, and emotional support.

We acknowledge that despite mobilizing numerous education experts and frontline teachers, the evaluation criteria for each scenario remain inadequate, with significant gaps in coverage. Consequently, we have shifted our focus to streamlining and automating the evaluation process, decoupling case design and assessment from complex evaluation systems. To facilitate large-scale, reproducible evaluations, we developed ELMES (Evaluation of Large Models in Educational Scenarios), an open-source, automated evaluation framework. Designed to address intricate pedagogical interactions beyond the reach of conventional methods, ELMES features a declarative workflow system. This allows researchers to effortlessly define and orchestrate dynamic, multi-agent dialogues (e.g., between a \textbf{teacher} and a \textbf{student}) through intuitive configuration files, eliminating the need for extensive coding. At its core, ELMES transforms traditionally subjective pedagogical metrics—such as instructional quality, guidance strategies, and emotional support—into structured, analyzable data through its modular hybrid evaluation engine (LLM-as-a-Judge). The framework provides a comprehensive toolchain that automates the entire workflow, from dialogue generation and process logging to multi-dimensional quantitative analysis. This ensures robust technical support for evaluation scenarios while maintaining efficiency, consistency, and reproducibility throughout the assessment process.

In summary, the main contributions of this paper are threefold:

\begin{itemize}
  \item We develop and open-source an automated evaluation framework, ELMES: Our work introduces ELMES, a modular and extensible framework that fundamentally decouples case-specific assessments from core evaluation methodology, enabling flexible adaptation across diverse educational scenarios. The framework automates the complete evaluation workflow - from dialogue generation and process logging to multi-dimensional quantitative analysis - significantly reducing manual effort while ensuring rigorous consistency and reproducibility. By open-sourcing this standardized infrastructure, we provide the research community with a scalable platform for systematic evaluation and improvement of educational LLMs.
  \item We propose four pedagogy-centered evaluation scenarios, task and metrics: Through iterative collaboration with domain experts, we have developed a comprehensive set of fine-grained, multi-dimensional evaluation metrics specifically designed for four critical educational scenarios: Knowledge Explanation, Guided Problem-Solving, Lesson Plan Generation, and Contextualized Question Posing. These metrics assess essential pedagogical qualities including adaptability, instructional clarity, creativity, and emotional intelligence, ensuring close alignment with authentic teaching requirements. The metrics underwent multiple refinement cycles incorporating expert feedback to guarantee both robustness and practical relevance in real-world educational settings.
  \item We conduct a systematic evaluation of state-of-the-art models: Leveraging the ELMES framework, we executed large-scale comparative analyses of multiple leading open-source and proprietary LLMs across various educational tasks. Our comprehensive evaluation reveals detailed capability distributions, highlighting both strengths and limitations of current models. These findings provide actionable insights for model developers to prioritize optimization efforts and practical guidance for educators in selecting appropriate AI tools. The results particularly illuminate critical gaps in current LLMs regarding personalization and contextual adaptation, offering clear directions for future research in educational AI.
\end{itemize}

The remainder of this paper is organized as follows: Section~\ref{sec:related_work} reviews related work, Section~\ref{sec:elmes_framework} details the ELMES framework, Section~\ref{sec:evaluation_scenarios} presents the evaluation scenarios, Section~\ref{sec:experiments} describes the experimental setup and results, and Section~\ref{sec:conclusion} concludes the paper with a discussion of future work.

\section{Related Work}
\label{sec:related_work}
The evolution of Large Language Model (LLM) evaluation benchmarks has progressed through several distinct phases. Initial benchmarks like GLUE~\cite{wang2018glue} and SuperGLUE~\cite{sarlin2020superglue} established foundational metrics for general language understanding. Subsequent developments, including MMLU~\cite{hendrycks2020measuring}, expanded evaluation scope to cross-disciplinary knowledge assessment. More recent benchmarks such as HelloBench~\cite{que2024hellobench} and LiveBench~\cite{white2024livebench} have introduced innovations in long-context generation and dynamic evaluation methodologies. While these benchmarks demonstrate increasing sophistication, they predominantly assess declarative knowledge ("what models know") rather than pedagogical competence ("how models teach") - a critical limitation for educational applications.

Recent efforts have begun addressing this gap through domain-specific educational benchmarks. Dr.Academy~\cite{chen2024dr} pioneered evaluating LLMs in educator roles through question generation tasks. Subject-specific evaluations have emerged, including MathBench's~\cite{liu2024mathbench} comprehensive mathematics framework and MATH2VISUAL's~\cite{wang2025generating} assessment of mathematical visualization capabilities. For advanced applications, GPQA~\cite{rein2024gpqa} and SuperGPQA~\cite{du2025supergpqa} focus on graduate-level problem-solving. However, these approaches remain constrained by single-turn, single-discipline evaluation paradigms that fail to capture the dynamic, multi-turn interactions characteristic of authentic teaching scenarios.

Our work represents a fundamental paradigm shift from previous approaches in three key aspects. First, rather than focusing solely on metric design, we prioritize evaluation process simplification through a modular framework. Second, while extensive interdisciplinary collaboration (AI+Education) informed our work, we recognize the inherent challenges in creating universally applicable assessment systems. Third, and most significantly, we introduce a configuration-based approach that decouples evaluation scenarios from core methodology. This innovation enables researchers to easily define domain-specific evaluation tasks through simple configuration files, transforming our framework into an extensible platform for pedagogical assessment. The resulting system supports dynamic, multi-turn teaching process simulation while maintaining rigorous evaluation standards - a substantial advance over existing static evaluation paradigms.

\section{The ELMES Framework}
\label{sec:elmes_framework}

\begin{figure*}[htbp]
    \centering
    \includegraphics[width=1\textwidth]{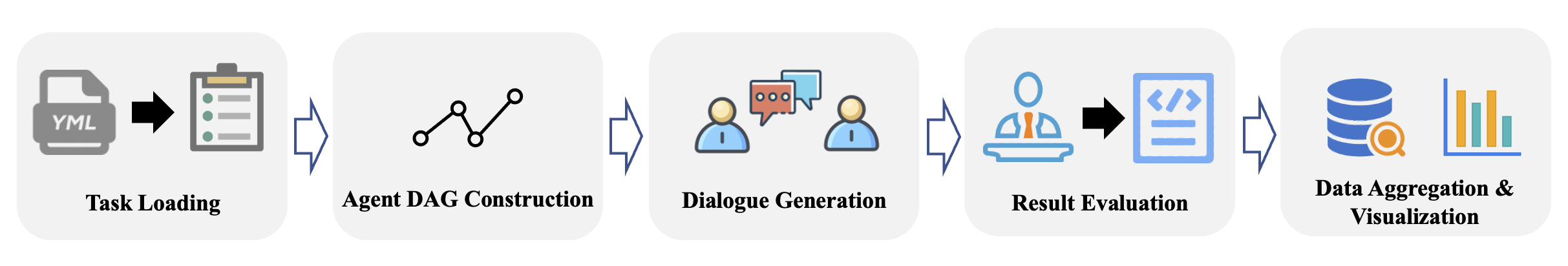}
    \caption{ELMES System Dataflow}
    \label{fig:elmes_workflow}
\end{figure*}

The rapid development of Large Language Models (LLMs) has opened up new possibilities in areas such as personalized education and intelligent tutoring. However, evaluating the true effectiveness of an LLM in educational contexts involves more than just measuring the accuracy of its knowledge base; it requires assessing its comprehensive capabilities in dynamic, multi-turn interaction scenarios like guided instruction and Socratic dialogue. Traditional evaluation benchmarks, often focused on single-turn Q\&A or static knowledge, are insufficient for assessing these complex tasks. To address this challenge, we introduce the ELMES framework, designed to solve a series of core problems. These include how to efficiently simulate complex teaching interactions involving multiple roles (e.g., teacher and student), how to flexibly orchestrate the conversational sequence and logical branching of agents, how to convert subjective metrics like teaching quality into quantifiable data, and how to ensure the complete traceability of large-scale experimental processes and results. This paper will detail the design and implementation of ELMES, showcasing its potential as a research infrastructure for educational large models.

\subsection{System Architecture and Design}

The architecture of ELMES deeply integrates four fundamental principles: modularity, extensibility, reproducibility, and automation. In terms of modularity, the core functions of the system are decoupled into independent Python modules, reducing dependencies between components and facilitating separate maintenance and extension. Its extensibility is manifested in a declarative configuration system entirely driven by YAML files, allowing users to adjust all aspects of an experiment without modifying the source code. To ensure reproducibility, all intermediate products of experiments, from dialogue history to evaluation results, are persistently stored, ensuring that any experiment can be precisely audited. Finally, through a suite of command-line tools covering the entire workflow, the framework achieves end-to-end automation, significantly enhancing research efficiency. 

The operation of ELMES follows a clear, multi-stage data pipeline, as illustrated in Figure~\ref{fig:elmes_workflow}. The process begins with the task loading stage, where the system parses a user-provided \textit{config.yaml} file and expands multiple sets of variables into a series of independent test cases based on a predefined mode. Subsequently, the framework enters the Agent DAG construction stage. Based on the \textit{directions} field in the configuration file, it instantiates all declared agents and constructs them into a Directed Acyclic Graph (DAG), where nodes represent agents and edges define the control flow transfer, supporting complex dialogue logic through a built-in routing mechanism. In the core dialogue generation stage, the system uses asynchronous mechanisms to concurrently process all test cases. The LangGraph engine orchestrates interactions between agents and the large language model according to the DAG, and all dialogues are saved in real-time to an SQLite database. After the dialogue is complete, the evaluation engine is activated to score the complete dialogue record against a predefined rubric. Finally, in the data aggregation and visualization stage, the framework's command-line tools can aggregate scattered evaluation results into a unified CSV report and generate multi-dimensional comparison charts with a single command, providing researchers with intuitive data analysis support.

\begin{tcolorbox}[
    title=Agent Field in ELMES Configuration File,
    label={tcb:agent},
    float*=ht!,
    width=\textwidth,
    breakable
]
\begin{verbatim}
agents:
  teacher:
    model: model1
    prompt:
      - role: system
        content: YOUR-TEACHER-SYSTEM-PROMPT-HERE
      - role: user
        content: "Here is the question for today’s one-on-one session: {question}"
    memory:
      keep_turns: 3
  student:
    model: model2
    prompt:
      - role: system
        content: |
          You are a student with the following profile: {image}.
          AND-YOUR-STUDENT-SYSTEM-PROMPT
\end{verbatim}
\end{tcolorbox}

\begin{tcolorbox}[title=Directions Field in ELMES Configuration File,
  float=ht!,
  label={tcb:directions},
  ]
\begin{verbatim}
directions:
  - START -> teacher
  - teacher -> router:any_keyword_route(
        keywords=[
          "class over", 
          "see you"
        ], 
        exists_to=END, 
        else_to="student"
    )
  - student -> teacher
\end{verbatim}
\end{tcolorbox}

\begin{tcolorbox}[title=Model Field in ELMES Configuration File,float=ht!,label={tcb:model}]
\begin{verbatim}
models:
  model1:
    type: openai
    api_key:
    model: gpt-4o-mini
  model2:
    ...
  model3:
    ...
\end{verbatim}
\end{tcolorbox}

\subsection{Core Mechanisms}

\subsubsection{Dynamic Scenarios and Declarative Workflows}

In ELMES, each experimental scenario is defined through a single, declarative configuration file. This file specifies the complete logic of the interaction, encompassing five core components that work in concert: \textbf{Agents}, \textbf{Models}, \textbf{Tasks}, \textbf{Directions}, and \textbf{Evaluation}.

\begin{itemize}
    \item \textbf{Agents} define the roles and behaviors of the participants in the simulation, such as a "teacher" or a "student". Each agent is configured with a specific language model and a system prompt that establishes its persona, knowledge base, and operational instructions.

    \item \textbf{Models} serve as a registry for the LLMs used in the experiment. This design allows for the seamless integration of any model compatible with the OpenAI API format, simply by adding a new configuration entry. This facilitates comparative studies across different LLMs.

    \item \textbf{Tasks} are central to generating dynamic and diverse scenarios. This section defines variables, such as different student personas or instructional questions, which are injected into agent prompts at runtime. This "configuration-as-code" approach allows researchers to systematically generate and test a vast number of scenarios, streamlining studies on the impact of specific variables on interaction outcomes.

    \item \textbf{Directions} specify the conversational workflow as a directed graph. It defines the sequence of interactions between agents (nodes) and includes support for conditional routing based on conversational cues. This enables the creation of complex workflows, from simple question-answer pairs to multi-turn debates with dynamic termination conditions.

    \item \textbf{Evaluation} automates the assessment of the interaction quality. It configures an LLM-based "judge" that analyzes the complete dialogue based on predefined criteria and outputs a structured assessment.
\end{itemize}

Collectively, this declarative, component-based architecture ensures that the entire experimental process is transparent, reproducible, and highly extensible. It abstracts away implementation details, allowing researchers to focus on the design and analysis of agent-based simulations.

\begin{tcolorbox}[title=Tasks Field in ELMES Configuration File,float=ht!,label={tcb:tasks}]
\begin{verbatim}
tasks:
  start_prompt:
    role: user
    content: ""
  mode: union
  content:
    image:
      - IMAGE1
      - IMAGE2
    question:
      - QUESTION1
      - QUESTION2
\end{verbatim}
\end{tcolorbox}

\begin{tcolorbox}[
    title=Evaluation Field in ELMES Configuration File,float=ht!,label={tcb:evaluation}
]
\begin{verbatim}
evaluation:
  model: model3
  name: config_modelname
  prompt:
    - role: system
      content: |
        YOUR-JUDGE-SYSTEM-PROMPT-HERE

        Indicators: 
        [Omitted for brevity in this excerpt]

        Exercise: 
        {task.question}

        Teaching Dialogue: 
        {messages.as_dialog()}

  format:
    - field: Accuracy
      type: int
      description: ...
    - field: Guidance
      type: int
      description: ...
    - field: Goal Alignment
      type: int
      description: ...
    - field: Personalization
      type: int
      description: ...
    - field: Metacognition
      type: int
      description: ...
    - field: Cultural Integration
      type: int
      description: ...
  format_mode: prompt
\end{verbatim}
\end{tcolorbox}

\subsubsection{Pluggable Hybrid Evaluation Engine}
Evaluation is a key component of ELMES, and its Metric Engine is designed for both flexibility and accuracy. The engine supports two types of configurable metrics: one is objective metrics, such as keyword matching and response length, which can be determined by simple rule-based functions. The other is more complex subjective metrics, such as the instructional quality and personalization of the teaching content. The assessment of these metrics is guided by the \textit{LLM-as-a-Judge} approach, which utilizes a powerful language model supplemented by a detailed scoring rubric.

All evaluation metrics are defined in a hierarchical structure within the configuration file. The framework automatically generates Pydantic data models based on this structure and provides two strategies for interacting with the evaluation model. The first is a Tool mode, which converts the data model into a tool that the LLM can call, ensuring the structured reliability of the output through function calls. The second is a Prompt mode, which injects the data model's JSON Schema and examples into the prompt, guiding the model to return formatted output within specific delimiters, offering broader compatibility.

\subsubsection{Automated Experiment Toolchain}
To achieve end-to-end automation, ELMES provides a complete suite of command-line tools covering the entire experiment lifecycle:

\begin{itemize}
    \item \textbf{\texttt{elmes generate}} – Concurrently executes test cases and generates dialogue data from the configuration file.
    \item \textbf{\texttt{elmes export}} – Exports data into standard formats for analysis or annotation.
    \item \textbf{\texttt{elmes eval}} – Calls the evaluation engine to score data and generate aggregated reports.
    \item \textbf{\texttt{elmes draw}} – Renders the agent workflow into an image.
    \item \textbf{\texttt{elmes visualize}} – Converts numerical reports into multi-dimensional comparison charts.
\end{itemize}

This toolchain standardizes experimental procedures into simple commands, enabling researchers to focus on strategy and performance analysis.

\subsection{Human Evaluation and LLM-as-a-Judge Integration}
In addition to automated scoring, ELMES integrates a robust \textbf{LLM-as-a-Judge} mechanism to evaluate the quality of model-generated outputs. Furthermore, it supports seamless incorporation of human judgment by enabling the export of evaluation data in formats compatible with the Label Studio annotation platform. Specifically, through the command \textit{elmes export label-studio}, ELMES can generate both the Label Interface configuration file (\textit{label-studio.txt}) and the corresponding data file (\textit{label-studio.json}). These files can be directly uploaded to Label Studio, allowing researchers to efficiently conduct human evaluations in parallel with or in comparison to LLM-based assessments. This dual evaluation strategy enhances the reliability, transparency, and interpretability of model performance in educational settings.

\section{Evaluation Scenarios}
\label{sec:evaluation_scenarios}
To systematically and multi-dimensionally evaluate the potential of large language models in education, we have designed four core evaluation scenarios: Knowledge Point Explanation, Guided Problem-Solving Instruction, Interdisciplinary Lesson Plan Generation, and Contextualized Problem Generation. As shown in Table~\ref{tab:scenarios_features}, further details will be introduced here.

\subsection{Knowledge Point Explanation}
This scenario aims to evaluate the model's ability to generate personalized explanations of knowledge points for specific learners. The core objective is to assess whether the model can move beyond generic descriptions to customize instructional content based on a predefined learner persona (e.g., grade level, cognitive ability, interests), thereby helping the learner build a clear and accurate conceptual understanding. To achieve this, our evaluation paradigm sets the model in a "teacher" role. It employs a single-turn interaction: the model receives a prompt containing a specific knowledge point, a teaching stage, and a learner persona, and must generate a complete explanatory text as the evaluation data. For the quantitative and qualitative assessment of this generated text, we have established an evaluation framework based on three primary dimensions: Role-Playing, Explanation Strategy, and Persona Responsiveness. The detailed evaluation metrics are shown in Table~\ref{knowledge_explanation}.

\subsection{Guided Problem-Solving Teaching}
This scenario is designed to evaluate the model's guidance capabilities in an interactive teaching context. It simulates a learner-centered, constructivist teaching method to assess whether the model can provide "scaffolding" through step-by-step questioning. This approach aims to guide learners to think independently and progressively construct a solution path, thereby cultivating their problem-solving skills. The evaluation employs a multi-turn dialogue paradigm simulating real student-teacher interactions: the model-under-test acts as the "teacher," interacting with a separate model playing the "student" based on a predefined persona, until the problem is solved or a preset turn limit is reached. To comprehensively assess its performance, we have developed an evaluation framework covering five primary dimensions: Reliability, Guided Instruction, Values, Creativity, and Emotional Support. The specific criteria are detailed in Table~\ref{guided_problem_solving_metrics_condensed}.

\subsection{Interdisciplinary Lesson Plan Generation}
This scenario aims to evaluate the model's ability to design interdisciplinary lesson plans. The core objective is to assess whether the model can organically integrate the knowledge, skills, and thinking methods from two or more disciplines around a central theme to create a highly integrative, cognitively progressive, and practical instructional plan. The evaluation paradigm is a single-turn generation task, where the model must generate a complete and structured interdisciplinary lesson plan based on a clear prompt containing a core theme, target disciplines, student grade level, and lesson duration. The evaluation of this plan is based on five primary dimensions: Theme and Objective Design, Activity and Task Design, Assessment System, Feasibility, and Logical Structure. The detailed metrics are shown in Table~\ref{interdisciplinary_design_metrics}.

\subsection{Contextualized Problem Generation}
This scenario aims to evaluate the model's ability to create contextualized math problems. This task requires the model not only to accurately grasp the knowledge point but also to seamlessly embed the assessment requirements into an authentic, engaging, and meaningful context. The goal is to evaluate its capacity to stimulate learner interest and assess the application of knowledge to real-world situations. The evaluation requires the model to generate a complete problem unit—including the question stem, answer, and solution—in a single turn based on a set of precise input parameters (knowledge point, core competencies, difficulty level, and format). We have constructed a comprehensive evaluation framework from five dimensions: Question Content Quality, Answer \& Solution Quality, Context Quality, Pedagogical Utility, and Value Alignment, as detailed in Table~\ref{contextualized_problem_gen_metrics}.

\section{Experiments}
\label{sec:experiments}
\subsection{Experimental Setup}
\textbf{Dataset.}
To systematically evaluate the comprehensive educational capabilities of AI models, we constructed an evaluation benchmark comprising four core dimensions. First, for Concept Explanation, we designed 10 cases pairing learner personas with specific knowledge points to assess the model's ability to deliver personalized instructional content. Second, in Guided Problem-Solving, we created 15 cases by crossing three student cognitive levels with five mathematical problems to evaluate the model's proficiency in guided teaching. Third, for Interdisciplinary Curriculum Design, we formulated 10 lesson plan generation tasks with complex constraints to examine the model's higher-order planning and innovation capabilities. Finally, for Contextualized Question Generation, we designed 10 tasks requiring the integration of specific knowledge points with core competencies to assess their capacity for creating high-quality, innovative assessment items. This benchmark's design spans from single-discipline applications to interdisciplinary integration in content, and from single-turn generation to multi-turn interaction in modality, thereby enabling a comprehensive and in-depth examination of the model's multifaceted capabilities across diverse educational scenarios.

\textbf{Metrics.} 
For our evaluation methodology, we adopt an automated evaluation paradigm based on a Large Language Model (LLM-as-a-Judge), selecting the \emph{gemini-2.5-pro-exp-03-25} model to serve as the evaluator. For each scenario, the evaluator model assigns a score from 1 to 5 based on predefined evaluation dimensions. This scoring adheres to a 5-point Likert scale~\cite{joshi2015likert}, with the following criteria: 
\begin{itemize}
  \item 1 : Completely fails to meet requirements
  \item 2 : Fails to meet requirements
  \item 3 : Meets requirements
  \item 4 : Exceeds requirements
  \item 5 : Completely exceeds requirements
\end{itemize}

Ultimately, we report two core types of metrics: first, the dimensional average score for each evaluation dimension across all cases, to assess the model's performance on specific capabilities independently; and second, an overall average score (Avg) that synthesizes all dimensions, to measure the model's comprehensive performance.

\textbf{Baselines.}
We compare our model with twelve closed-source and open-source models, as listed below:

\begin{itemize}
    \item \textbf{GPT-4o~\cite{hurst2024gpt}:} A native model that processes and generates arbitrary combinations of text, audio, and images end-to-end through a single neural network.
    
    \item \textbf{Qwen-2.5-32B-Instruct~\cite{team2024qwen2}:} A representative large-parameter open-source model from the Qwen2.5 series, which benefits from the series' comprehensive pre-training and post-training techniques while maintaining high efficiency.
    
    \item \textbf{Qwen-2.5-72B-Instruct~\cite{team2024qwen2}:} The flagship open-source instruction-tuned model of the Qwen2.5 series, which significantly enhances its instruction-following and human preference alignment capabilities through expanded pre-training data and multi-stage post-training.
    
    \item \textbf{DeepSeek-V3~\cite{liu2024deepseek}:} A powerful Mixture-of-Experts (MoE) language model with 671B total parameters, which leverages an efficient architecture to achieve performance comparable to leading closed-source models.
    
    \item \textbf{DeepSeek-R1~\cite{guo2025deepseek}:} A reasoning-focused model trained with large-scale reinforcement learning to elicit strong emergent reasoning capabilities.
    
    \item \textbf{QwQ-32B~\cite{QwenTeam2025QWQ}:} A 32B-parameter reasoning model that achieves performance comparable to much larger models by applying large-scale reinforcement learning on a strong base model, and integrates agent-like capabilities.
    
    \item \textbf{Spark-v4.0-Ultra~\cite{iFlytekSpark2024}:} A large language model that demonstrates leading performance on a range of tasks, including text generation, language understanding, and logical reasoning.

    \item \textbf{Qwen3-235B-A22B~\cite{yang2025qwen3}:} An innovative open-source model series that, at its core, integrates a "thinking mode" for complex reasoning with a "non-thinking mode" for swift responses, enabling users to dynamically manage the trade-off between performance and latency through a "thinking budget" mechanism.

    \item \textbf{Claude-Opus-4-20250514~\cite{anthropic2025claude}:} The state-of-the-art model, which sets a new world-leading standard for complex, agentic coding tasks, bringing frontier-level performance to everyday developer and enterprise applications.

    \item \textbf{Gemini-2.5-Pro-Preview-06-05~\cite{comanici2025gemini}:} The state-of-the-art model, optimizing the trade-off between performance and efficiency for advanced AI tasks.

    \item \textbf{Grok-4~\cite{xai2025grok}:} The world's most intelligent model, featuring native tool use and real-time search, whose advanced reasoning is achieved by scaling reinforcement learning to a massive new level with vast compute and data.
    
    \item \textbf{MuduoLLM~\cite{MuduoLLM2025}:} A domain-specific large model designed for K-12 education, focusing on educational scenario capabilities such as guided Q\&A and lesson plan generation.
\end{itemize}

\subsection{Results and Analysis}
Our evaluation of the knowledge point explanation task reveals a significant performance trade-off between pedagogical delivery and content mastery (see Table \ref{tab:knowledge_point_explanation}). Notably, Gemini-2.5-Pro-Preview-06-05 (AVG: 4.55) achieves the highest overall score, excelling not in static knowledge recall but in dynamic instructional metrics such as Appropriateness of Teaching Methods (TM: 4.30) and Response to Personalization (RP: 4.70). This performance profile contrasts with the two distinct paradigms exemplified by Qwen-2.5-32B-Instruct (AVG: 4.28), which demonstrates superior Emotional Support (ES: 5.00), and DeepSeek-V3-0324 (AVG: 4.22), which excels in foundational Knowledge Mastery (KM: 5.00).
Furthermore, the results suggest that increased model scale does not uniformly improve performance; both Qwen-2.5-72B-Instruct (AVG: 4.13) and Qwen3-235B-A22B (AVG: 3.87) underperform relative to smaller, more specialized models in key instructional dimensions. The most salient evidence of capability decoupling is presented by DeepSeek-R1-0528 (AVG: 3.98), whose high Knowledge Mastery (KM: 4.90) and low Role Adherence (RA: 2.15) scores indicate that factual accuracy and interactive persona may be independent, and at times conflicting, model attributes. Models such as GPT-4o-2024-11-20 (AVG: 4.03) further support this, exhibiting high Role Adherence (RA: 4.35) but a pronounced weakness in Knowledge Mastery (KM: 3.55). Finally, the limited capabilities of models like Spark-v4.0-Ultra (AVG: 2.78) reinforce that effective instructional dialogue is a specialized function requiring targeted optimization, not an inherent property of language models.

\begin{table}[htbp]
\centering
\caption{
    A comprehensive evaluation of models on their ability to explain knowledge points. 
    This table assesses models across 6 fine-grained metrics, with the highest score in each column highlighted in \textbf{bold}.
    The abbreviated metrics are as follows: 
    \textbf{RA} (Role Adherence), 
    \textbf{ES} (Emotional Support), 
    \textbf{KM} (Knowledge Mastery), 
    \textbf{TM} (Appropriateness of Teaching Methods), 
    \textbf{CD} (Appropriateness of Content Design), and
    \textbf{RP} (Response to Personalization).
}
\label{tab:knowledge_point_explanation}
\footnotesize
\setlength{\tabcolsep}{3.5pt} 
\begin{tabular}{@{}lccccccc@{}}
\toprule
\textbf{Model} & \textbf{RA} & \textbf{ES} & \textbf{KM} & \textbf{TM} & \textbf{CD} & \textbf{RP} & \textbf{AVG} \\
\midrule
GPT-4o-2024-11-20         & \textbf{4.35} & 4.70          & 3.55          & 3.60          & 3.95          & 4.00          & 4.03 \\
Qwen-2.5-32B-Instruct     & \textbf{4.35} & \textbf{5.00} & 4.20          & 4.20          & 4.80          & 3.15          & 4.28 \\
Qwen-2.5-72B-Instruct     & 3.55          & 4.80          & 3.80          & 4.00          & 4.25          & 4.35          & 4.13 \\
DeepSeek-V3-0324          & 2.95          & 4.80          & \textbf{5.00} & 4.20          & \textbf{4.95} & 3.40          & 4.22 \\
DeepSeek-R1-0528          & 2.15          & 4.75          & 4.90          & 4.15          & 4.90          & 3.05          & 3.98 \\
QwQ-32B                   & 2.90          & 4.75          & 4.45          & 4.25          & 4.80          & 2.50          & 3.94 \\
Spark-v4.0-Ultra          & 2.40          & 2.65          & 3.25          & 3.05          & 3.00          & 2.35          & 2.78 \\
Qwen3-235B-A22B           & 2.55          & 4.40          & 3.95          & 3.50          & 4.55          & 4.25          & 3.87 \\
Claude-Opus-4-20250514    & 3.65          & 4.35          & 4.40          & 4.10          & 4.35          & 4.40          & 4.21 \\
Gemini-2.5-Pro-Preview-06-05 & 3.90      & 4.95          & 4.65          & \textbf{4.30} & 4.80          & \textbf{4.70} & \textbf{4.55} \\
Grok-4                    & 3.90          & 4.85          & 3.80          & 3.50          & 4.30          & 3.85          & 4.04 \\
MuduoLLM                  & 2.60          & 2.50          & 4.05          & 3.55          & 3.60          & 2.80          & 3.19 \\
\bottomrule
\end{tabular}
\end{table}

\begin{table*}[htbp]
\centering
\caption{
    Comprehensive evaluation of models on guided problem-solving capabilities. 
    This table assesses models across 14 fine-grained metrics, with the highest score in each column highlighted in \textbf{bold}.
    The abbreviated metrics are as follows: 
    \textbf{RC} (Role Consistency), 
    \textbf{TR} (Topic Relevance), 
    \textbf{AC} (Accuracy), 
    \textbf{GU} (Guidance), 
    \textbf{DI} (Directionality), 
    \textbf{PE} (Personalization), 
    \textbf{MC} (Metacognition Cultivation), 
    \textbf{CI} (Cultural Integration), 
    \textbf{VO} (Value Orientation), 
    \textbf{CT} (Critical Thinking), 
    \textbf{DT} (Divergent Thinking), 
    \textbf{EI} (Emotional Insight), 
    \textbf{EA} (Emotional Appropriateness), and 
    \textbf{IT} (Inclusivity and Trust).
}
\label{tab:guided_problem_solving_teaching}
\resizebox{\textwidth}{!}{%
\begin{tabular}{@{}lccccccccccccccc@{}}
\toprule
\textbf{Model} & \textbf{RC} & \textbf{TR} & \textbf{AC} & \textbf{GU} & \textbf{DI} & \textbf{PE} & \textbf{MC} & \textbf{CI} & \textbf{VO} & \textbf{CT} & \textbf{DT} & \textbf{EI} & \textbf{EA} & \textbf{IT} & \textbf{AVG} \\
\midrule
GPT-4o-2024-11-20         & 4.53          & 4.67          & 4.13          & 4.53          & 4.40          & 4.33          & 4.07          & 2.80          & 3.33          & 3.73          & 4.40          & 4.40          & 4.60          & 4.60          & 4.18 \\
Qwen-2.5-32B-Instruct     & \textbf{5.00} & \textbf{4.73} & 4.47          & 4.53          & 4.27          & 4.53          & 3.67          & 2.13          & 3.33          & 3.40          & 3.60          & 4.07          & 4.73          & 4.67          & 4.08 \\
Qwen-2.5-72B-Instruct     & \textbf{5.00} & 4.53          & 4.67          & 4.67          & 4.33          & 4.73          & 4.33          & 2.53          & 3.47          & 4.20          & 4.00          & 4.73          & 4.60          & 4.67          & 4.32 \\
DeepSeek-V3-0324          & 3.67          & 3.67          & 3.67          & 3.67          & 3.67          & 3.53          & 3.20          & 2.27          & 2.53          & 3.13          & 2.93          & 3.40          & 3.60          & 3.60          & 3.32 \\
DeepSeek-R1-0528          & 2.20          & 2.27          & 2.07          & 2.33          & 2.13          & 2.27          & 2.13          & 1.73          & 1.60          & 2.00          & 1.87          & 2.27          & 2.33          & 2.27          & 2.10 \\
QwQ-32B                   & 2.27          & 2.13          & 2.33          & 2.33          & 2.07          & 2.33          & 2.13          & 1.27          & 1.60          & 1.80          & 1.73          & 2.20          & 2.13          & 2.33          & 2.05 \\
Spark-v4.0-Ultra          & 4.73          & \textbf{4.73} & 4.53          & 4.67          & \textbf{4.73} & 4.73          & 4.53          & 1.47          & 3.20          & 4.00          & 3.67          & 4.47          & 4.73          & 4.60          & 4.20 \\
Qwen3-235B-A22B           & 4.33          & 1.47          & 3.20          & 3.00          & 2.33          & 3.20          & 2.13          & 2.07          & 2.00          & 2.00          & 2.20          & 2.33          & 3.40          & 3.13          & 2.63 \\
Claude-Opus-4-20250514    & 4.80          & 3.07          & 4.73          & 4.93          & 3.00          & 4.93          & \textbf{4.73} & 3.67          & 4.07          & 4.33          & 4.27          & 4.80          & \textbf{4.93} & \textbf{5.00} & \textbf{4.38} \\
Gemini-2.5-Pro-Preview-06-05 & 4.87          & 2.27          & 4.33          & 4.93          & 3.60          & \textbf{5.00} & 4.33          & 4.40          & \textbf{4.60} & 4.07          & 4.27          & 4.87          & 4.87          & 4.93          & \textbf{4.38} \\
Grok-4                    & \textbf{5.00} & 1.53          & \textbf{4.87} & \textbf{5.00} & 2.47          & \textbf{5.00} & 4.53          & \textbf{4.47} & 4.33          & \textbf{4.60} & \textbf{4.80} & \textbf{4.93} & 4.67          & \textbf{5.00} & 4.37 \\
MuduoLLM                  & 2.60          & 1.13          & 1.93          & 2.40          & 2.13          & 2.20          & 1.80          & 1.07          & 1.47          & 1.73          & 1.53          & 2.20          & 2.40          & 2.27          & 1.92 \\
\bottomrule
\end{tabular}%
}
\end{table*}

The evaluation of Guided Problem-Solving Teaching (Table \ref{tab:guided_problem_solving_teaching}) delineates clear performance tiers and reveals that leading models exhibit distinct, specialized pedagogical profiles rather than uniform superiority. Claude-Opus-4 (AVG: 4.38) and Gemini-2.5-Pro (AVG: 4.38) co-lead, yet demonstrate this divergence: Claude excels in affective and metacognitive support, leading in Metacognition Cultivation (MC: 4.73) and Emotional Appropriateness (EA: 4.93), whereas Gemini excels in adaptive instruction, securing top scores in Personalization (PE: 5.00) and Value Orientation (VO: 4.60). Closely following, Grok-4 (AVG: 4.37) establishes a third profile focused on cognitive scaffolding, dominating metrics like Accuracy (AC: 4.87), Critical Thinking (CT: 4.60), and Divergent Thinking (DT: 4.80).
A critical trade-off emerges from these results: top-performing models often show deficiencies in foundational metrics like Topic Relevance (TR) and Directionality (DI). This suggests that advanced creative and adaptive capabilities may currently come at the cost of maintaining strict pedagogical focus, a dimension where a model like Spark-v4.0-Ultra (AVG: 4.20) excels (DI: 4.73). The performance of other models, including the consistent Qwen-2.5-72B (AVG: 4.32), further populates this complex capability spectrum. Conversely, the significant performance drop-off for models such as Qwen3-235B-A22B (AVG: 2.63) and DeepSeek-R1-0528 (AVG: 2.10) underscores the task's inherent complexity. These results indicate that effective multi-turn instructional guidance is a highly specialized function requiring targeted optimization.

The Interdisciplinary Lesson Plan Generation task reveals a clear bifurcation in model capabilities, distinguishing between models optimized for deep structural design and those excelling at formal coherence (Table \ref{tab:lesson_plan_generation}). The DeepSeek series demonstrates a significant advantage in this highly structured task, with DeepSeek-R1-0528 (AVG: 4.50) achieving the highest overall score. Its top performance in core metrics such as Interdisciplinary Logical Integration (IL: 5.00) and Cognitive Progression Design (CP: 4.95) validates its strength in complex pedagogical reasoning. The strong performance of QwQ-32B (AVG: 4.43), which notably leads in Cognitive Conflict Design (CF: 2.90), further underscores the superiority of specialized models in crafting advanced instructional elements.
In stark contrast, GPT-4o (AVG: 4.17) exemplifies a critical trade-off. While it achieves perfect scores in formal attributes like Lesson Plan Standardization (LS: 5.00) and Internal Consistency (IC: 5.00), it scores lowest of all models in the pedagogically crucial Cognitive Conflict Design (CF: 1.60). This disparity suggests a risk of generating outputs that are structurally polished but lack substantive pedagogical depth. This profile of formal strength over pedagogical substance is distinct from that of other general-purpose models. Models such as Spark-v4.0-Ultra (AVG: 4.13), Gemini-2.5-Pro (AVG: 4.25), and Claude-Opus-4 (AVG: 3.94) occupy a middle performance tier, producing lesson plans that are generally coherent and usable but fail to match the innovative design of the top-performing specialized models. Finally, the significant performance drop-off for models like Grok-4 (AVG: 3.66) and MuduoLLM (AVG: 3.42) reinforces the high technical barrier of this task, confirming that sophisticated lesson plan generation is not an emergent property but a specialized capability requiring targeted optimization.
\begin{table*}[htbp]
\centering
\caption{
    Comprehensive evaluation of models on interdisciplinary lesson plan generation. 
    This table assesses models across 15 fine-grained metrics, with the highest score in each column highlighted in \textbf{bold}.
    The abbreviated metrics are as follows: 
    \textbf{CC} (Core Concept Connectivity), 
    \textbf{IL} (Interdisciplinary Logical Integration), 
    \textbf{KC} (Core Knowledge Coverage), 
    \textbf{CF} (Cognitive Conflict Design), 
    \textbf{CA} (Context Authenticity), 
    \textbf{CP} (Cognitive Progression Design), 
    \textbf{DS} (Differentiated Support), 
    \textbf{SE} (Student Engagement), 
    \textbf{PR} (Process Reflection), 
    \textbf{CE} (Closed-Loop Evaluation Design), 
    \textbf{ME} (Multi-dimensional Performance Evaluation), 
    \textbf{FC} (Feasibility of Conditions), 
    \textbf{MT} (Material and Tool Appropriateness), 
    \textbf{LS} (Lesson Plan Standardization), and 
    \textbf{IC} (Internal Consistency).
}
\label{tab:lesson_plan_generation}
\resizebox{\textwidth}{!}{%
\setlength{\tabcolsep}{4pt} 
\begin{tabular}{@{}lcccccccccccccccc@{}} 
\toprule
\textbf{Model} & \textbf{CC} & \textbf{IL} & \textbf{KC} & \textbf{CF} & \textbf{CA} & \textbf{CP} & \textbf{DS} & \textbf{SE} & \textbf{PR} & \textbf{CE} & \textbf{ME} & \textbf{FC} & \textbf{MT} & \textbf{LS} & \textbf{IC} & \textbf{AVG} \\
\midrule
GPT-4o-2024-11-20         & 4.80          & 4.75          & 4.90          & 1.60          & 3.80          & 4.65          & 2.60          & \textbf{5.00} & 2.80          & 4.70          & \textbf{5.00} & 3.90          & 4.10          & \textbf{5.00} & \textbf{5.00} & 4.17 \\
Qwen-2.5-32B-Instruct     & 4.70          & 4.70          & 4.90          & 1.40          & 4.25          & 4.35          & 2.80          & \textbf{5.00} & 3.20          & 4.60          & 4.70          & 3.30          & 4.10          & 4.90          & \textbf{5.00} & 4.13 \\
Qwen-2.5-72B-Instruct     & 3.70          & 4.00          & 4.70          & 2.10          & 4.10          & 4.30          & 2.00          & 4.90          & 2.60          & 4.20          & 4.80          & 3.40          & 4.00          & 4.80          & 4.50          & 3.87 \\
DeepSeek-V3-0324          & \textbf{5.00} & 4.90          & \textbf{5.00} & 2.30          & 4.80          & 4.80          & 3.55          & \textbf{5.00} & 2.30          & 4.75          & \textbf{5.00} & 3.90          & 4.70          & \textbf{5.00} & \textbf{5.00} & 4.40 \\
DeepSeek-R1-0528          & 4.90          & \textbf{5.00} & \textbf{5.00} & 2.70          & \textbf{4.90} & \textbf{4.95} & \textbf{3.60} & \textbf{5.00} & 3.40          & \textbf{4.85} & 4.90          & 3.45          & \textbf{4.80} & \textbf{5.00} & \textbf{5.00} & \textbf{4.50} \\
QwQ-32B                   & \textbf{5.00} & 4.90          & 4.50          & \textbf{2.90} & 4.60          & 4.70          & 2.70          & \textbf{5.00} & \textbf{3.80} & 4.80          & \textbf{5.00} & \textbf{4.00} & 4.70          & \textbf{5.00} & 4.90          & 4.43 \\
Spark-v4.0-Ultra          & 4.50          & 4.50          & 4.60          & 2.20          & 4.40          & 4.60          & 2.50          & 4.90          & 2.80          & 4.20          & 4.70          & 3.60          & 4.50          & 4.90          & \textbf{5.00} & 4.13 \\
Qwen3-235B-A22B           & 4.44          & 4.56          & 4.22          & 2.33          & 4.22          & 4.00          & 2.56          & 4.44          & 3.11          & 3.67          & 4.33          & 3.78          & 4.22          & 4.22          & 4.78          & 3.93 \\
Claude-Opus-4-20250514    & 4.60          & 4.40          & 4.20          & 2.50          & 3.90          & 3.90          & 2.50          & 4.20          & 3.10          & 4.10          & 4.30          & \textbf{4.00} & 4.30          & 4.30          & 4.80          & 3.94 \\
Gemini-2.5-Pro-Preview-06-05 & 4.60          & 4.80          & 4.70          & 2.70          & 4.80          & 4.20          & 2.90          & 4.80          & 3.30          & 4.40          & 4.50          & \textbf{4.00} & 4.70          & 4.40          & \textbf{5.00} & 4.25 \\
Grok-4                    & 3.80          & 4.00          & 4.10          & 2.20          & 4.00          & 3.60          & 2.10          & 4.00          & 3.00          & 3.80          & 4.00          & 3.50          & 4.20          & 3.80          & 4.80          & 3.66 \\
MuduoLLM                  & 4.20          & 3.20          & 3.90          & 2.40          & 4.20          & 3.70          & 2.00          & 4.10          & 2.40          & 2.60          & 2.90          & 3.90          & 3.90          & 3.70          & 4.20          & 3.42 \\
\bottomrule
\end{tabular}%
}
\end{table*}

Evaluation of the Contextualized Problem Generation task reveals a clear hierarchy of models, distinguishing between well-rounded leaders and those with more specialized or limited capabilities (Table \ref{tab:problem_generation}). Gemini-2.5-Pro (AVG: 4.46) emerges as the top-performing model, demonstrating a balanced profile of excellence with a leading score in Solution Quality (SQ: 4.5) and consistently high marks across all other metrics. This establishes it as a robust all-rounder.
Closely trailing are several models exhibiting distinct strengths. Claude-Opus-4 (AVG: 4.34) and Grok-4 (AVG: 4.34) are tied in overall score, yet present different profiles. Claude excels in the pedagogical aspects of the task, securing top scores in Pedagogical Utility (PU: 5.0) and Context Quality (CQ: 4.8), though this comes with a notable trade-off in Solution Quality (SQ: 3.1). Grok-4, alongside Qwen3-235B-A22B (AVG: 4.42), provides consistently strong, well-rounded performance, solidifying the top competitive tier. In a specialist capacity, QwQ-32B (AVG: 3.96) distinguishes itself by co-leading in Problem Quality (PQ: 4.7), indicating a particular strength in the final generation output, even if its overall performance is lower.
In contrast, a number of other prominent models, including GPT-4o (AVG: 3.32) and Spark-v4.0-Ultra (AVG: 3.46), form a distinct lower-performance tier. The notably low score of GPT-4o in Solution Quality (SQ: 2.0), for instance, highlights a critical deficiency. The performance gap between these models and the leaders underscores the task's complexity, suggesting that high-quality contextualized problem generation demands an integrated and highly optimized synthesis of comprehension, pedagogical design, and generation quality.
\begin{table}[htbp]
\centering
\caption{
    Evaluation of models on contextualized problem generation. 
    The highest score in each column is highlighted in \textbf{bold}.
    The abbreviated metrics are: 
    \textbf{VA} (Value Alignment), 
    \textbf{CQ} (Context Quality), 
    \textbf{PU} (Pedagogical Utility), 
    \textbf{SQ} (Solution Quality), and 
    \textbf{PQ} (Problem Quality).
}
\label{tab:problem_generation}
\footnotesize
\begin{tabular}{@{}lcccccr@{}}
\toprule
\textbf{Model} & \textbf{VA} & \textbf{CQ} & \textbf{PU} & \textbf{SQ} & \textbf{PQ} & \textbf{AVG} \\
\midrule
GPT-4o-2024-11-20         & 3.9 & 3.8 & 3.6 & 2.0 & 3.3 & 3.32 \\
Qwen-2.5-32B-Instruct     & 3.6 & 2.5 & 3.6 & 3.7 & 3.3 & 3.34 \\
Qwen-2.5-72B-Instruct     & 4.0 & 4.3 & 4.3 & 3.9 & 4.1 & 4.12 \\
DeepSeek-V3-0324          & 4.0 & 3.2 & 3.7 & 3.3 & 3.7 & 3.58 \\
DeepSeek-R1-0528          & 3.4 & 2.7 & 4.4 & 3.8 & 4.0 & 3.66 \\
QwQ-32B                   & 3.4 & 3.3 & 4.3 & 4.1 & \textbf{4.7} & 3.96 \\
Spark-v4.0-Ultra          & 3.8 & 3.4 & 3.7 & 2.8 & 3.6 & 3.46 \\
Qwen3-235B-A22B           & \textbf{4.6} & 4.7 & 4.7 & 3.8 & 4.3 & 4.42 \\
Claude-Opus-4-20250514    & 4.1 & \textbf{4.8} & \textbf{5.0} & 3.1 & \textbf{4.7} & 4.34 \\
Gemini-2.5-Pro-Preview-06-05 & 4.2 & 4.3 & 4.8 & \textbf{4.5} & 4.5 & \textbf{4.46} \\
Grok-4                    & 3.8 & 4.4 & 4.9 & 4.1 & 4.5 & 4.34 \\
MuduoLLM                  & 3.5 & 3.1 & 2.5 & 2.1 & 2.9 & 2.82 \\
\bottomrule
\end{tabular}%
\end{table}

\begin{figure}[ht]
    \centering
    \includegraphics[width=0.5\textwidth]{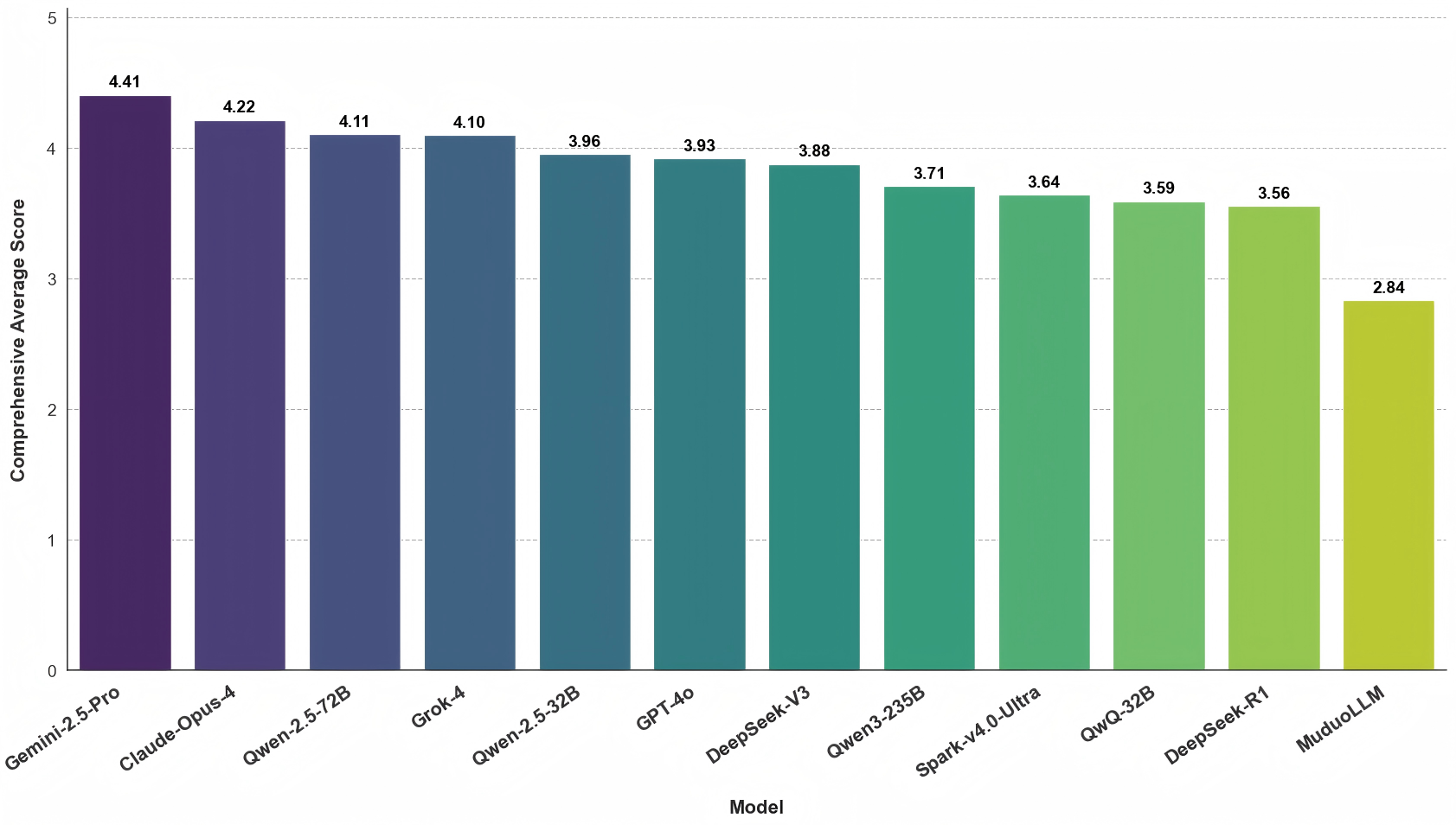}
    \caption{Comprehensive Performance Comparison of Twelve Large Language Models. The chart displays the comprehensive average score for each model across all evaluated tasks, revealing a clear performance hierarchy.}
    \label{fig:overall}
\end{figure}

As shown in Figure \ref{fig:overall}, analysis of the comprehensive average scores reveals a distinct, tiered performance landscape among the twelve evaluated models. Gemini-2.5-Pro establishes a clear performance apex at 4.41, followed by a tightly clustered elite cohort comprising Claude-Opus-4 (4.22), Qwen-2.5-72B (4.11), and Grok-4 (4.10), all scoring above 4.1. A subsequent competitive tier shows marginal differentiation, with Qwen-2.5-32B (3.96), GPT-4o (3.93), and DeepSeek-V3 (3.88) grouped closely together. The performance then steps down to a broader mid-tier, including Qwen3-235B (3.71), Spark-v4.0-Ultra (3.64), QwQ-32B (3.59), and DeepSeek-R1 (3.56). Notably, MuduoLLM (2.84) represents a significant outlier, demonstrating a marked performance gap relative to all other models. However, this distribution critically reveals that even the most advanced general-purpose models fail to achieve perfect proficiency, underscoring a persistent gap between their broad capabilities and the nuanced, specialized requirements inherent to high-quality educational tasks.

\section{Conclusion}
\label{sec:conclusion}
Current benchmarks for Large Language Models (LLMs) assess general intelligence but fail to measure the pedagogical capabilities essential for education. To address this critical gap, we introduced a comprehensive evaluation system centered on four authentic educational scenarios and developed ELMES, an open-source framework for its automated and reproducible implementation. Our systematic evaluation of state-of-the-art LLMs revealed that different models exhibit distinct pedagogical profiles, often displaying a trade-off between deep content mastery and effective interactive teaching. These findings provide crucial empirical guidance for developers to optimize models for educational tasks and for educators to select tools aligned with specific learning objectives.
Our primary contributions—the pedagogy-centered evaluation system and the ELMES framework—establish a new paradigm for assessing the true educational utility of LLMs. While foundational, this work invites future research to expand the scope of evaluation scenarios and to continue refining the LLM-as-a-Judge methodology to ensure its robustness. By shifting the focus from general capabilities to specific pedagogical effectiveness, this research paves the way for the development of more capable, responsible, and impactful AI-powered educational tools.


\bibliographystyle{elsarticle-num} 
\bibliography{bare_jrnl.bib}

\newpage
\appendix

\section{Scenario Evaluation Metrics}

\subsection{Knowledge Point Explanation}
This table~\ref{knowledge_explanation} outlines metrics for evaluating an AI's teaching performance when explaining a knowledge point. The assessment focuses on three core dimensions: first, Role-Playing, requiring the AI to emulate a competent primary school teacher in its language, emotional support, and knowledge. Second is Teaching Strategy, which assesses the use of effective methods like guided questioning and relatable real-life examples. Finally, the rubric measures the AI's ability to provide Personalized responses based on the student's individual information, such as grade or interests.

\begin{table*}[htbp]
\centering
\caption{Evaluation Metrics for Knowledge Point Explanation}
\label{knowledge_explanation}
\begin{tabular}{>{\bfseries}l p{4cm} p{7cm}}
\toprule
\textbf{Primary Dimension} & \textbf{Secondary Metric} & \textbf{Description} \\
\midrule
\textbf{Role-Playing} & 
Role Adherence & 
1. Strictly follows the persona of a primary school teacher, using a first-person perspective. \newline 
2. Acknowledges its nature as an AI model by not attempting physical interactions or describing its own expressions and actions, and does not explain its instructional design. \newline 
3. Demonstrates the knowledge base, linguistic style, and characteristics appropriate for a primary school math teacher. \\

& 
Emotional Support & 
Provides guidance, affirmation, and emotional support. \\

& 
Knowledge Mastery & 
Exhibits a thorough understanding of the user-specified knowledge point. \\ \midrule

\textbf{Explanation Strategy} & 
Appropriateness of Teaching Methods & 
1. Poses questions to elicit the student's viewpoint at appropriate times; often guides the student to think before providing an answer. \newline 
2. Encourages student self-reflection on the learned content. \\

& 
Appropriateness of Content Design & 
Uses examples within the cognitive range of primary school students. Focuses on applying math to solve real-life problems with authentic scenarios. \\ \midrule

\textbf{Persona Responsiveness} & 
Response to Personalization & 
Responds to the student's persona, such as their grade, age, or expressed interests, and other personal information. \\

\bottomrule
\end{tabular}
\end{table*}

\subsection{Guided Problem-Solving Teaching}

This table~\ref{guided_problem_solving_metrics_condensed} details the metrics for evaluating the effectiveness of guided problem-solving teaching. The core of the evaluation lies in the "guided" nature of the teaching, which emphasizes inspiring students through questioning rather than direct instruction, and adapting the approach based on individual student needs. Additionally, it considers multiple dimensions such as content accuracy, the cultivation of creative thinking, emotional support, and even the integration of cultural values, aiming for a holistic educational experience.

\begin{table*}[htbp]
\centering
\caption{Evaluation Metrics for Guided Problem-Solving Teaching}
\label{guided_problem_solving_metrics_condensed}
\begin{tabular}{>{\bfseries}l p{4.5cm} p{8cm}}
\toprule
\textbf{Primary Dimension} & \textbf{Secondary Metric} & \textbf{Description} \\
\midrule

\textbf{Content Fidelity} & 
Accuracy & 
Evaluates the mathematical accuracy and rigor of all presented knowledge (concepts, formulas) and the entire problem-solving process (reasoning, calculation, notation). \\
&
Role Consistency & 
Evaluates if the model consistently adheres to its preset role in language, knowledge, and thinking style, avoiding persona breaks. \\
&
Topic Relevance &
Evaluates if the dialogue stays focused on the core topic, avoiding irrelevant digressions. \\

\midrule

\textbf{Guided} & 
Guidance & 
Evaluates the use of progressive questions to guide student thinking toward self-discovery, rather than providing direct answers. \\
&
Directionality &
Evaluates if the model maintains focus on the learning objective, guiding exploration toward an effective solution. \\
&
Personalization &
Evaluates the ability to adapt instruction (depth, pace, style) based on the student's specific knowledge, understanding, and error patterns. \\
&
Metacognition Cultivation &
Evaluates if the model prompts students to reflect on their problem-solving process (e.g., reviewing steps, analyzing errors) to improve self-monitoring. \\

\midrule

\textbf{Values} &
Cultural Integration &
Evaluates the integration of China's mathematical heritage, philosophy, and social context into the learning content and tasks. \\
&
Value Orientation &
Evaluates if the model fosters social responsibility via real-world applications and connects mathematical rigor to virtues like integrity. \\

\midrule

\textbf{Creativity} &
Critical Thinking &
Evaluates if the model encourages students to question information, identify assumptions, and analyze problems from multiple perspectives. \\
&
Divergent Thinking &
Evaluates if the model encourages exploring multiple solutions, provides open-ended problem-solving space, and supports creative approaches. \\

\midrule

\textbf{Emotional Support} &
Emotional Insight &
Evaluates the ability to identify specific student emotions (e.g., frustration, confusion) and respond with targeted, genuine empathy. \\
&
Emotional Appropriateness &
Evaluates if emotional support (e.g., encouragement, humor) is timely, well-calibrated, and serves the learning objective without being distracting. \\
&
Inclusivity and Trust &
Evaluates if the model fosters a safe, positive, and mistake-tolerant environment that encourages questions and exploration. \\

\bottomrule
\end{tabular}
\end{table*}

\subsection{Interdisciplinary Lesson Plan Generation}

This table~\ref{interdisciplinary_design_metrics} presents a comprehensive set of evaluation metrics for interdisciplinary lesson plan design. It primarily assesses whether the plan logically integrates knowledge from multiple disciplines and stimulates student thinking through authentic, scaffolded activities. Furthermore, it examines the completeness of the evaluation system, the practical feasibility of the plan, and the overall logical coherence of its structure.

\begin{table*}[htbp]
\centering
\caption{Evaluation Metrics for Interdisciplinary Lesson Plan Generation}
\label{interdisciplinary_design_metrics}
\begin{tabular}{>{\bfseries}l p{4.5cm} p{8cm}}
\toprule
\textbf{Primary Dimension} & \textbf{Secondary Metric} & \textbf{Description} \\
\midrule

\textbf{Theme and Objective Design} & 
Core Concept Connectivity & 
Evaluates the clarity and logical coherence of connections between core concepts, ensuring they form a structured knowledge network. \\
& 
Interdisciplinary Logical Integration & 
Evaluates the logical integration of multiple disciplines to solve authentic, real-world problems, rather than a simple patchwork of topics. \\
&
Core Knowledge Coverage & 
Evaluates if the lesson covers essential, foundational knowledge points (e.g., 3+ key concepts or methods) for the topic. \\

\midrule

\textbf{Activity and Task Design} & 
Cognitive Conflict Design & 
Evaluates the intentional design of cognitive conflicts (e.g., counter-intuitive results) and guided inquiry to resolve them and build correct understanding. \\
&
Context Authenticity & 
Evaluates if learning scenarios are authentic, relatable to students' lives, and use real-world data or problems to foster engagement. \\
& 
Cognitive Progression Design & 
Evaluates if the lesson plan systematically scaffolds learning from foundational knowledge to higher-order thinking skills. \\
&
Differentiated Support & 
Evaluates if the plan provides differentiated support and challenges (e.g., in tasks, resources) to meet diverse learner needs. \\
&
Student Engagement & 
Evaluates if the design promotes active student participation (e.g., discussion, collaboration) over passive information reception. \\

\midrule

\textbf{Evaluation System Design} & 
Process Reflection & 
Evaluates if the plan includes reflection activities for students to record and analyze their learning process, promoting self-regulation. \\
&
Closed-Loop Evaluation Design & 
Evaluates the design of a complete feedback loop (formative and summative) where assessment aligns with goals and informs learning. \\
&
Multi-dimensional Performance Evaluation & 
Evaluates if assessment is multi-dimensional (e.g., knowledge, skills, thinking) and based on clear, observable performance indicators. \\

\midrule

\textbf{Implementation Feasibility} & 
Feasibility of Conditions & 
Evaluates the practical feasibility regarding time, materials, and equipment within a standard school environment. \\
&
Material and Tool Appropriateness & 
Evaluates if materials and tools are appropriate for the target students' age and cognitive level and effectively support learning objectives. \\

\midrule

\textbf{Logical Structure} &
Lesson Plan Standardization &
Evaluates the formal structure (e.g., formatting, numbering) and the professional clarity, accuracy, and conciseness of the language used. \\
&
Internal Consistency & 
Evaluates the logical coherence and alignment between all components of the lesson plan, such as objectives, activities, and assessments. \\

\bottomrule
\end{tabular}
\end{table*}

\subsection{Contextualized Problem Generation}

This table~\ref{contextualized_problem_gen_metrics} provides metrics for evaluating the quality of generated contextualized problems. It focuses not only on the fundamental quality of the problem content and its solution, such as accuracy and clarity, but also places a strong emphasis on the design of the context, requiring it to be authentic, appropriate, and tightly integrated with the knowledge points. Furthermore, the rubric specifically examines the "pedagogical utility" of the problem, meaning its potential as a teaching tool to inspire student thinking, and its underlying "value alignment."

\begin{table*}[htbp]
\centering
\caption{Evaluation Metrics for Generated Contextualized Problem Generation}
\label{contextualized_problem_gen_metrics}
\begin{tabular}{>{\bfseries}l p{11.5cm}}
\toprule
\textbf{Primary Dimension} & \textbf{Description} \\
\midrule

Problem Content Quality &
Evaluates if the problem statement is clear, solvable (with a unique answer if single-choice), and accurately assesses the specified knowledge points and core competencies at the designated difficulty level. \\

\midrule

Answer and Solution Quality &
Evaluates if the answer is accurate and the solution provides a clear, logical, step-by-step explanation using mainstream methods relevant to the topic, without introducing out-of-scope knowledge. \\

\midrule

Problem Context Quality &
Evaluates if the context is authentic, cognitively appropriate, and seamlessly integrated with the knowledge points. The scenario should be clearly described and align with positive societal and educational values. \\

\midrule

Pedagogical Utility &
Evaluates the potential for the problem and solution to be used as a teaching aid. This includes its ability to stimulate student thinking, facilitate classroom discussion, and support teaching activities like concept clarification or error analysis. \\

\midrule

Value Alignment &
Evaluates alignment with positive societal values (e.g., honesty, diligence) while strictly avoiding negative content (e.g., violence, stereotypes). Problems with explicit positive guidance should be rated highly. This metric focuses only on the value dimension. \\

\bottomrule
\end{tabular}
\end{table*}










\end{document}